\documentclass[12pt,a4paper]{./styles/iopart}
\usepackage[american]{babel}
\usepackage{rotating}
\usepackage{epsfig}
\usepackage{amsmath}
\usepackage{amssymb}
\usepackage{multirow}
\usepackage{axodraw}
\usepackage{dcolumn}
\usepackage{./styles/myxspace}
\usepackage{calc}
\usepackage{./styles/mcite}
\usepackage{cite}
\def\Euro{\hbox{\kern0.15em C\kern-0.7em%
\raisebox{-0.1ex}{--}\kern-0.43em\raisebox{0.15ex}{--}\kern0.13em}}

\def\Z0{{Z^\zero}}
\def\text#1{{\rm #1}}
\def\eVdist{\kern-0.06667em}

\def\Gev{{\text{Ge}\eVdist\text{V\/}}}
\def\Tev{{\text{Te}\eVdist\text{V\/}}}

\def\gev{{\,\text{Ge}\eVdist\text{V\/}}}

\def\pb{\,\text{pb}}

\def\pbi{\,\text{pb}^{-1}}

\def\BR{{\rm BR}}

\def\DO{{D{\O}}\xspace}

\def\IP{{\rm I$\kern-0.01667em$P}\xspace}

\def\LQsub{{\rm LQ}}
\def\LQ{{\rm LQ}}

\def\Ptlj{{\not{\kern-0.55ex P}}_t\ell j}
\def\Ptmiss{{\not{\kern-0.55ex P}}_t}

\def\lsqp#1{{\lambda'_{#1}}}

\def\ptmiss{{\not{\kern-0.3em P_t}}}

\mathchardef\qsm=63
\mathchardef\pls=43
\mathchardef\mns=512
\mathchardef\plm=518
\mathchardef\eql=61
\mathchardef\smallleft=300
\mathchardef\smallright=301
\mathchardef\perslsh=47
\mathchardef\les=316
\mathchardef\gre=318
\mathchardef\leq=532
\mathchardef\grq=533
\chardef\usc=95
\chardef\til=126

\def\sqr#1#2#3{{\vcenter{\hrule height.#3ex\hbox{\vrule width.#2ex height#1ex
    \kern#1ex\vrule width.#3ex}\hrule height.#2ex}}}

\def\angleto{\vrule width.035em height2.1ex depth-.56ex\unskip\kern-.6ex\to}
\def\perchc#1{{\raise.4ex\hbox{$\mkern4mu#1{\it\perslsh}_
             {\mkern-5mu\scriptscriptstyle{{\rm o}\!{\rm o}}}^
             {\mkern-12.8mu\scriptscriptstyle{\rm o}}$}}}

\def\widebar#1{\mkern1.5mu\overline{\mkern-1.5mu#1\mkern-1.mu}\mkern1.mu}
\catcode`\@=11 
\def\parenbar{\mathpalette\p@renb@r}
\def\p@renb@r#1#2{\vbox{%
  \ifx#1\scriptscriptstyle \dimen@.7em\dimen@ii.2em\else
  \ifx#1\scriptstyle \dimen@.8em\dimen@ii.25em\else
  \dimen@1em\dimen@ii.4em\fi\fi \offinterlineskip
  \ialign{\hfill##\hfill\cr
    \vbox{\hrule width\dimen@ii}\cr
    \noalign{\vskip-.3ex}%
    \hbox to\dimen@{$\mathchar300\hfil\mathchar301$}\cr
    \noalign{\vskip-.3ex}%
    $#1#2$\cr}}}
\catcode`\@=12 
\def\nuan{\parenbar{\nu}}

\def\pbar{\widebar{p}}

\def\qbar{\widebar{q}}

\def\nubar{\widebar{\nu}}

\newbox\struttbox
\setbox\struttbox=\hbox{\vrule height1.65ex depth.485ex width0pt}
\def\strutt{\relax\ifmmode\copy\struttbox\else\unhcopy\struttbox\fi}
\def\stru#1#2{\relax\ifmmode\hbox{\vrule height#1 depth#2 width0pt}
\else\vrule height#1 depth#2 width0pt\fi}

\def\ronum#1{\uppercase\expandafter{\romannumeral#1}}
\def\ronuml#1{\expandafter{\romannumeral#1}}

\def\cA{{\phantom{0}}}

\def\cbk{\kern-0.5em}
\newcommand{\pcite}[1]{{\protect\cite{#1}}}

\def\fig#1{Fig.~\ref{fig-#1}}
\def\Fig#1{Figure~\ref{fig-#1}}
\def\figand#1#2{Figs.~\ref{fig-#1} and~\ref{fig-#2}}
\def\Figand#1#2{Figures~\ref{fig-#1} and~\ref{fig-#2}}
\def\tab#1{Table~\ref{tab-#1}}

\def\sec#1{Sect.~\ref{sec-#1}}

\DeclareMathAlphabet{\mathbf}{OT1}{cmr}{bx}{sl}

\catcode`\@=11 
\let\tab@penalty\relax
\newcount\tab@state
\def\bcline#1{%
  \noalign{\kern-.5\arrayrulewidth\tab@penalty}%
  \omit%
  \global\tab@state\@ne%
  \ranges\bcline@i{#1}%
  \cr%
  \noalign{\kern-.5\arrayrulewidth\tab@penalty}%
}
\def\bcline@i#1#2{%
  \ifnum#1<\tab@state\relax%
    \tab@@cr%
    \noalign{\kern-\arrayrulewidth\tab@penalty}%
    \omit%
    \global\tab@state\@ne%
  \fi%
  \@whilenum\tab@state<#1\do{%
    \hfil\tab@@tab@omit%
    \global\advance\tab@state\@ne%
  }%
  \ifnum\tab@state>\@ne%
    \kern-\arrayrulewidth%
  \fi%
  \@whilenum\tab@state<#2\do{%
    \tab@@span@omit%
    \global\advance\tab@state\@ne%
  }%
  \leaders\hrule\@height\boldarrayrulewidth\hfill%
}
\def\ranges#1#2{%
  \gdef\ranges@temp{#1}%
  \begingroup%
  \ranges@i#2 \q@delim%
}
\def\ranges@i{%
  \@ifnextchar\q@delim\ranges@done{\afterassignment\ranges@ii\count@}%
}
\def\ranges@ii{%
  \@ifnextchar-\ranges@iii{\ranges@do\count@\count@\ranges@v}%
}
\def\ranges@iii-{\afterassignment\ranges@iv\@tempcnta}
\def\ranges@iv{\ranges@do\count@\@tempcnta\ranges@v}
\def\ranges@v{%
  \@ifnextchar,%
    \ranges@vi%
    {%
      \@ifnextchar\q@delim%
        \ranges@done%
        {\tab@err@range\ranges@vi,}%
    }%
}
\def\ranges@vi,{\afterassignment\ranges@ii\count@}
\def\ranges@do#1#2{%
  \ifnum#1>#2\else%
    \expandafter\endgroup%
    \expandafter\ranges@temp%
    \expandafter{%
    \the\expandafter#1%
    \expandafter}%
    \expandafter{%
    \the#2%
    }%
    \begingroup%
  \fi%
}
\def\ranges@done\q@delim{\endgroup}
\def\ifinrange#1#2{%
  \@tempswafalse%
  \count@#1%
  \ranges\ifinrange@i{#2}%
  \if@tempswa%
    \expandafter\@firstoftwo%
  \else%
    \expandafter\@secondoftwo%
  \fi%
}
\def\ifinrange@i#1#2{%
  \ifnum\count@<#1 \else\ifnum\count@>#2 \else\@tempswatrue\fi\fi%
}
\def\tab@@cr{\cr}
\def\tab@@tab@omit{&\omit}
\def\tab@@span@omit{\span\omit}
\def\tab@checkrule#1{%
  \count@#1\relax%
  \expandafter\ifinrange%
  \expandafter\count@%
  \expandafter{\tab@xcols}%
    {\tab@checkrule@i}%
    {}%
}
\def\bhline{\noalign{\ifnum0=`}\fi\hrule \@height  
\boldarrayrulewidth \futurelet \@tempa\@xhline}
\def\@xhline{\ifx\@tempa\hline\vskip \doublerulesep\fi
      \ifnum0=`{\fi}}
\catcode`\@=12 
\newcommand{\topboldline}{\bhline\bs}

\catcode`\@=11 
\newcounter{pict@width}
\newcounter{pict@height}
\newlength{\pict@scale}
\newcommand{\psfigadd}[4]{%
\setcounter{pict@width}{1*\ratio{#2+\pict@scale/2}{\pict@scale}}
\setcounter{pict@height}{1*\ratio{#3+\pict@scale/2}{\pict@scale}}
\setlength{\unitlength}{\pict@scale}
\hbox{\begin{picture}(\thepict@width,\thepict@height)
\put(0,0){\psfig{figure=#1,width=#2,height=#3,clip=}}
\SetScale{0.283466457}
\SetWidth{1.763889}
{#4}
\end{picture}}
}
\newcommand{\psfigror}[4]{%
\setcounter{pict@width}{1*\ratio{#2+\pict@scale/2}{\pict@scale}}
\setcounter{pict@height}{1*\ratio{#3+\pict@scale/2}{\pict@scale}}
\setlength{\unitlength}{\pict@scale}
\hbox{\begin{picture}(\thepict@width,\thepict@height)
\put(0,\thepict@height){\psfig{figure=#1,width=#3,height=#2,clip=,angle=270}}
\SetScale{0.283466457}
\SetWidth{1.763889}
{#4}
\end{picture}}
}
\newcommand{\psfigrol}[4]{%
\setcounter{pict@width}{1*\ratio{#2+\pict@scale/2}{\pict@scale}}
\setcounter{pict@height}{1*\ratio{#3+\pict@scale/2}{\pict@scale}}
\setlength{\unitlength}{\pict@scale}
\hbox{\begin{picture}(\thepict@width,\thepict@height)
\put(0,0){\psfig{figure=#1,width=#3,height=#2,clip=,angle=90}}
\SetScale{0.283466457}
\SetWidth{1.763889}
{#4}
\end{picture}}
}
\def\thebibliography#1{\list
 {\hfil[\arabic{enumi}]}{\topsep=0\p@\parsep=0\p@
 \partopsep=0\p@\itemsep=0\p@
 \labelsep=5\p@\itemindent=-0\p@
 \settowidth\labelwidth{\footnotesize[#1]}%
 \leftmargin\labelwidth
 \advance\leftmargin\labelsep
 \advance\leftmargin -\itemindent
 \usecounter{enumi}}\footnotesize
 \def\newblock{\ }
 \sloppy\clubpenalty4000\widowpenalty4000
 \sfcode`\.=1000\relax}
\catcode`\@=12 
\catcode`\@=12 

\usepackage{./styles/mysidecaption}
\renewcommand\baselinestretch{0.833}
\begin{document}
\selectlanguage{american}
\rightline{{\small Presented at
                   {\it Beyond the Desert 2002}, Oulu, Finland, 2--7 June 2002}
           \hfill
           {\bf FAU-PI1-02-01}}
\rightline{\bf 18~Dec.~2002}

\title[Searches for leptoquarks and excited fermions at HERA]%
      {Searches for leptoquarks\\ and excited fermions at HERA}

\author{Ulrich F. Katz\footnote{$\!\!$supported by a grant by the German Federal
                                Ministry for Education and Research}%
                      \footnote{$\!\!$on leave of absence from 
                                Physics Institute, University of Bonn} (ZEUS)\\
        {\small Representing the H1 and ZEUS Collaborations}}

\address{University of Erlangen-Nuremberg, Institute of Physics,\\ 
         Erwin-Rommel-Str.\ 1, 91058 Erlangen, Germany\\
         Email: katz@physik.uni-erlangen.de}

\begin{abstract}
Recent results on searches for new particles at the electron-proton collider
HERA are reported. Based on roughly $100\pbi$ of $e^+p$ data and $16\pbi$ of
$e^-p$ data per experiment, taken in the years 1994--2000, the H1 and ZEUS
collaborations have derived new exclusion limits for the direct production of
excited fermion states and leptoquarks. The latter are searched for in different
decay channels, including lepton-flavor violating decays. The production of
$R_P$-violating squarks followed by leptoquark-like decays to lepton and quark
is studied, as are cascade decays yielding multi-jet plus lepton signatures. New
limits from indirect searches are also reported. Several of the searches obtain
sensitivities of the same order or exceeding those of other experiments,
indicating the substantial discovery potential of future HERA running.
\end{abstract}

\newcommand{\citeZEUSes}{%
\cite{zeus-01-08,*zeus-budapest-607,zeus-exoweb}\xspace}

\section{Introduction}
\label{sec-int}

\looseness=-1
From 1998 to 2000, the electron-proton collider HERA at DESY was operated with
electrons or positrons at an energy of $E_e=27.5\gev$ and protons of
$E_p=920\gev$, yielding a center-of-mass energy of $\sqrt{s}=318\gev$. In this
period, the ZEUS and H1 collaborations collected $e^+p$ ($e^-p$) data samples
corresponding to integrated luminosities of about $66\pbi$ ($16\pbi$) per
experiment. Combining these data sets with previous $e^+p$ data at
$\sqrt{s}=300\gev$ increases the HERA sensitivity in searches for new heavy
particles to masses of almost $300\gev$ and production cross sections as low as
$0.1\pb$.

This document describes the status of searches for signals of excited fermions,
leptoquarks and $R_P$-violating squarks. It focuses on results that include the
data taken since 1998 and are thus the ``last word'' from HERA before the
high-luminosity phase HERA-2, which is currently beginning. Earlier results have
e.g.\ been reported at the {\sc Beyond99} conference \cite{bsm99}. Many of the
results presented in the following are still preliminary, and some have only
become available since the {\sc Beyond02} conference but are included to provide
an up-to-date overview of recent experimental results. A recent review of
searches at HERA and at the Tevatron is also available in \cite{kuz-02-01}.

\section{Excited fermions}
\label{sec-efer}

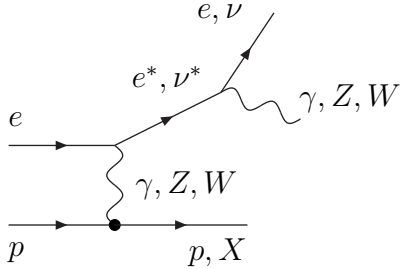
\begin{figure}[t]
  \sidecaption
  \begin{picture}(170.,90.)(0.,0.)
  \SetOffset(0.,-20.)
  \ArrowLine(10,30)(50,30)            
  \ArrowLine(10,60)(50,60)            
  \Photon(50,30)(50,60){3}{2}         
  \GCirc(50,30){2}{0}                 
  \ArrowLine(50,30)(100,30)           
  \ArrowLine(50,60)(90,80)            
  \ArrowLine(90,80)(110,110)          
  \Photon(90,80)(120,70){3}{2}        
  \Text(10,70)[l]{$e$}                
  \Text(10,20)[l]{$p$}                
  \Text(100,20)[r]{$p,X$}             
  \Text(58,45)[l]{$\gamma,Z,W$}       
  \Text(70,80)[b]{$e^\ast,\nu^\ast$}  
  \Text(90,110)[]{$e,\nu$}            
  \Text(120,78)[l]{$\gamma,Z,W$}      
  \end{picture}
  \caption{Feynman graph of the production and subsequent decay of excited
           electrons or neutrinos in $ep$ reactions. In the case of $e^\ast$
           production, elastic reactions $ep\to e^\ast p$ are possible.}
           \label{fig-es-fey}
\end{figure}

If electrons or neutrinos are composite, their excitations could be produced
in $ep$ reactions at HERA by $t$-channel exchange of photons, $Z$ or $W$ bosons
as shown in the diagram of \fig{es-fey}.  The observation of such states would
be a clear indication of fermion substructure.

\begin{table}[t]
  \caption{Decay modes, event signatures and main SM background sources  
           considered in the $e^\ast$ and $\nu^\ast$  searches by H1 
           \pcite{h1-02-01,h1-02-02} and ZEUS 
           \pcite{zeus-01-08,*zeus-budapest-607}.}
  \centerline{
  \begin{tabular}{llll}
  \topboldline
  $f^\ast$ decay&signature&main SM background&studied by\\
  \hline
  $e^\ast\to e+\gamma$ 
    &$e+\gamma$               &QEDC, NC DIS                     & H1, ZEUS\\
  $e^\ast\to e+Z\to e+q\qbar$
    &$e+2$jets                &NC DIS                           & H1\\
  $e^\ast\to\nu+W\to\nu+q\qbar'$ 
    &$\ptmiss+2$jets          &CC DIS, PHP                      & H1\\
  \hline
  $\nu^\ast\to\nu+\gamma$ 
    &$\gamma+\ptmiss$         &CC DIS                           & H1, ZEUS\\
  $\nu^\ast\to\nu+Z\to\nu+q\qbar$ 
    &$\ptmiss+2$jets          &CC DIS, PHP                      & H1, ZEUS\\
  $\nu^\ast\to e+W\to e+q\qbar'$
    &$e+2$jets                &NC DIS                           & H1, ZEUS\\
  \hline
  \end{tabular}}
  \label{tab-es-decmod}
\end{table}

The H1 \cite{h1-02-01,h1-02-02} and ZEUS \citeZEUSes collaborations have
searched for excited electrons ($e^\ast$) and neutrinos ($\nu^\ast$) under the
assumption that they decay into standard fermions and electroweak gauge bosons
in one of the modes summarized in \tab{es-decmod}. Whereas $e^\ast$ production
is expected to occur in $e^+p$ and $e^-p$ reactions with about the same cross
section, $\nu^\ast$ would be produced in $e^-p$ scattering at a much higher rate
than in $e^+p$ reactions, so the $\nu^\ast$ searches focus on the $e^-p$
data. The main selection cuts for both the $e^\ast$ and $\nu^\ast$ searches
required events consistent with the presence of both a final-state lepton with
high transverse momentum, $P_t$ (either an electron identified in the central
detector or large missing transverse momentum, $\ptmiss$, indicating a neutrino)
and of the gauge boson from the $f^\ast$ decay (either identified directly in
the case of a photon, or via the hadronic decay products in the case of $W$ and
$Z$ bosons). The backgrounds from various Standard Model (SM) processes (mainly
neutral current (NC) and charged current (CC) deep-inelastic scattering (DIS),
QED Compton scattering (QEDC) and photoproduction of jets with high transverse
momentum (PHP), see \tab{es-decmod}) are estimated from MC simulations. The
signal reactions are simulated using cross sections calculated according to the
model by Hagiwara, Zeppenfeld and Komamiya (HZK) \cite{hag-85-01} for spin-$1/2$
excited fermions.

\begin{figure}[t]
  \sidecaption
  \psfig{file=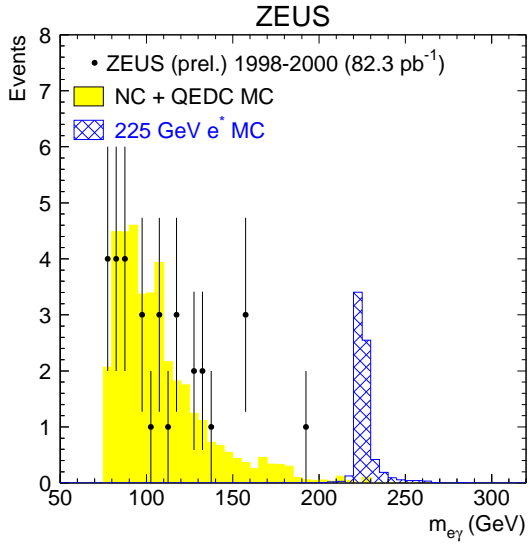,width=6.9cm}\kern-2.mm
  \caption{Spectrum of the invariant $e\gamma$ mass of ZEUS candidates for
           $e^\ast$ production and subsequent decay $e^\ast\to e\gamma$. The
           points with error bars show the data and the light-shaded histogram
           the total SM background. The hatched histogram illustrates the signal
           expected for a hypothetical $e^\ast$ with a mass of $225\gev$.}
           \label{fig-es-egam}
\end{figure}

\begin{figure}[b]
  \centerline{\strut\kern5.mm
  \psfig{file=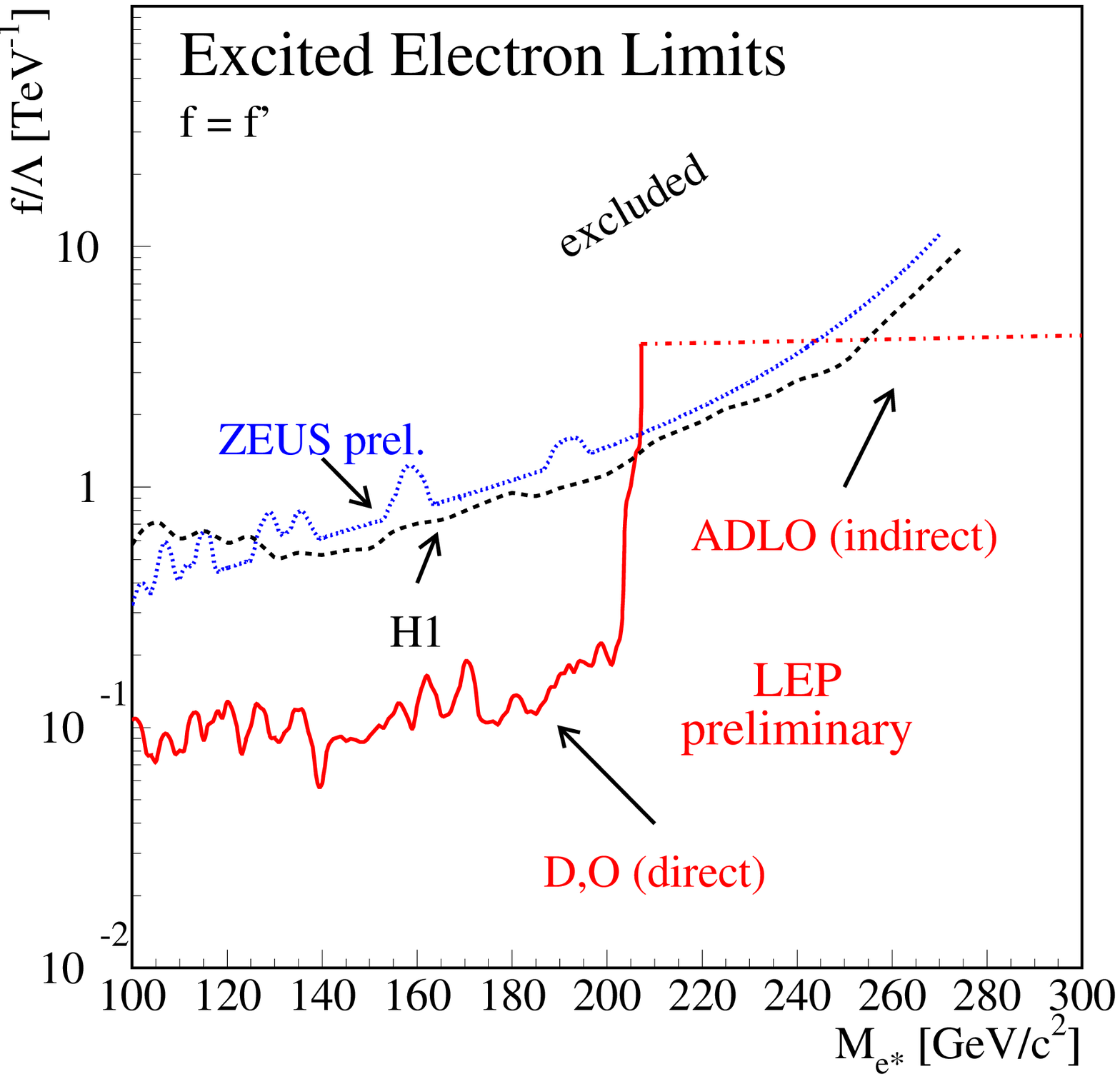,width=7.5cm,height=8.cm}
  \psfig{file=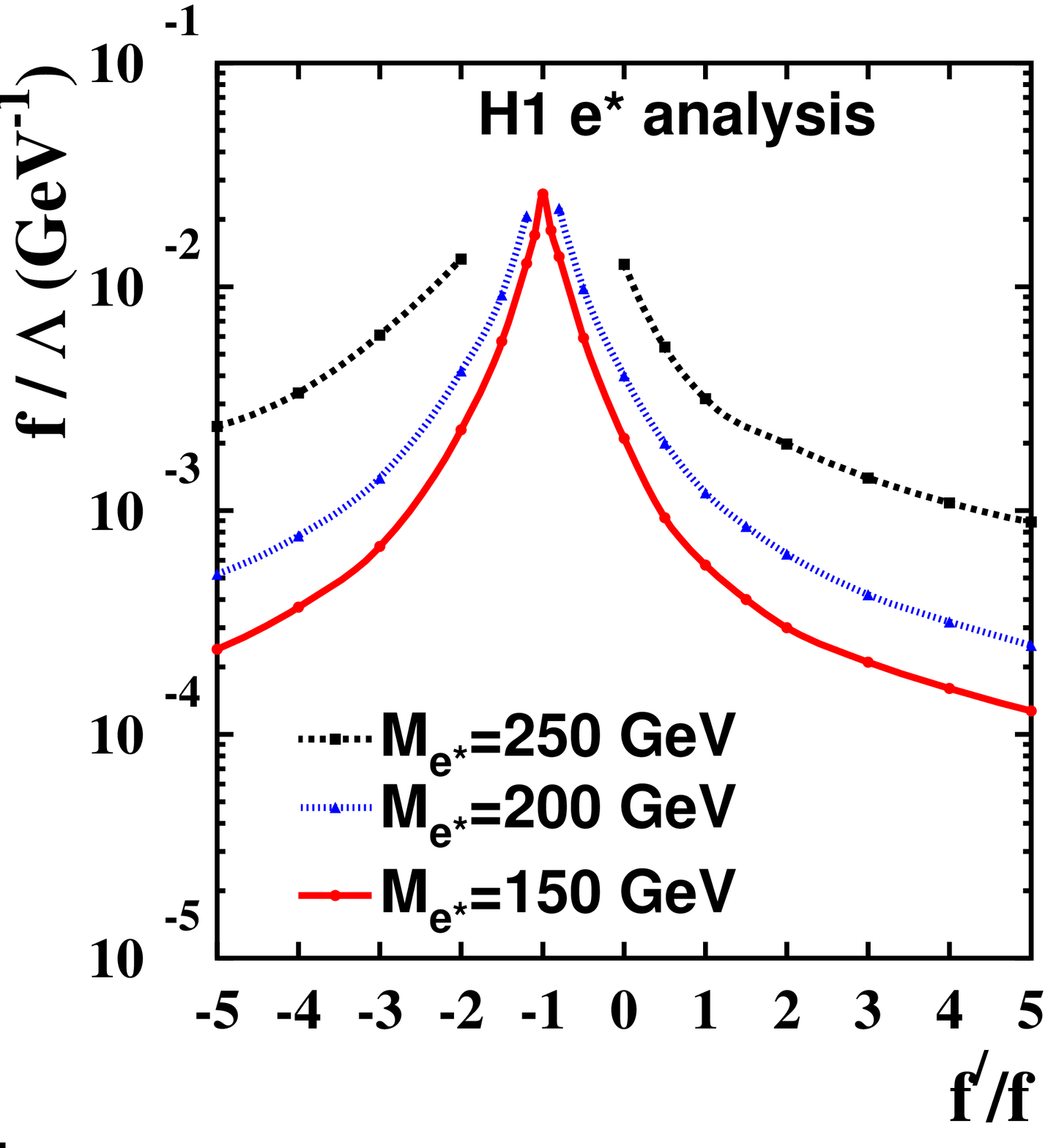,width=7.5cm,height=8.3cm}}
  \caption{Left: limits at $95\%$ C.L.\ on $f/\Lambda$ from the H1 and ZEUS
           searches for excited electrons compared to corresponding results
           from the LEP experiments. Right: dependence of the H1 limits on the
           ratio $f'/f$.}
  \label{fig-es-flam}
\end{figure}

No indication of a signal above SM background has been found. As an example,
\fig{es-egam} shows the distribution of the invariant $e\gamma$ mass in the
event sample selected in the ZEUS search for $e^\ast\to e\gamma$ decays,
together with the expected signal shape for an $e^\ast$ with
$M_{e^\ast}=225\gev$ (arbitrary normalization). For excited fermion ($f^\ast$)
masses $M_{f^\ast}\gtrsim100\gev$ the searches are almost background-free,
yielding upper limits at $95\%$ C.L.\ on the cross section times branching ratio
($\sigma\cdot\BR$) of typically $0.05\mns0.1\pb$ for the decay channels with
highest sensitivity. 

\begin{figure}[t]
  \centerline{\strut\kern5.mm
  \psfig{file=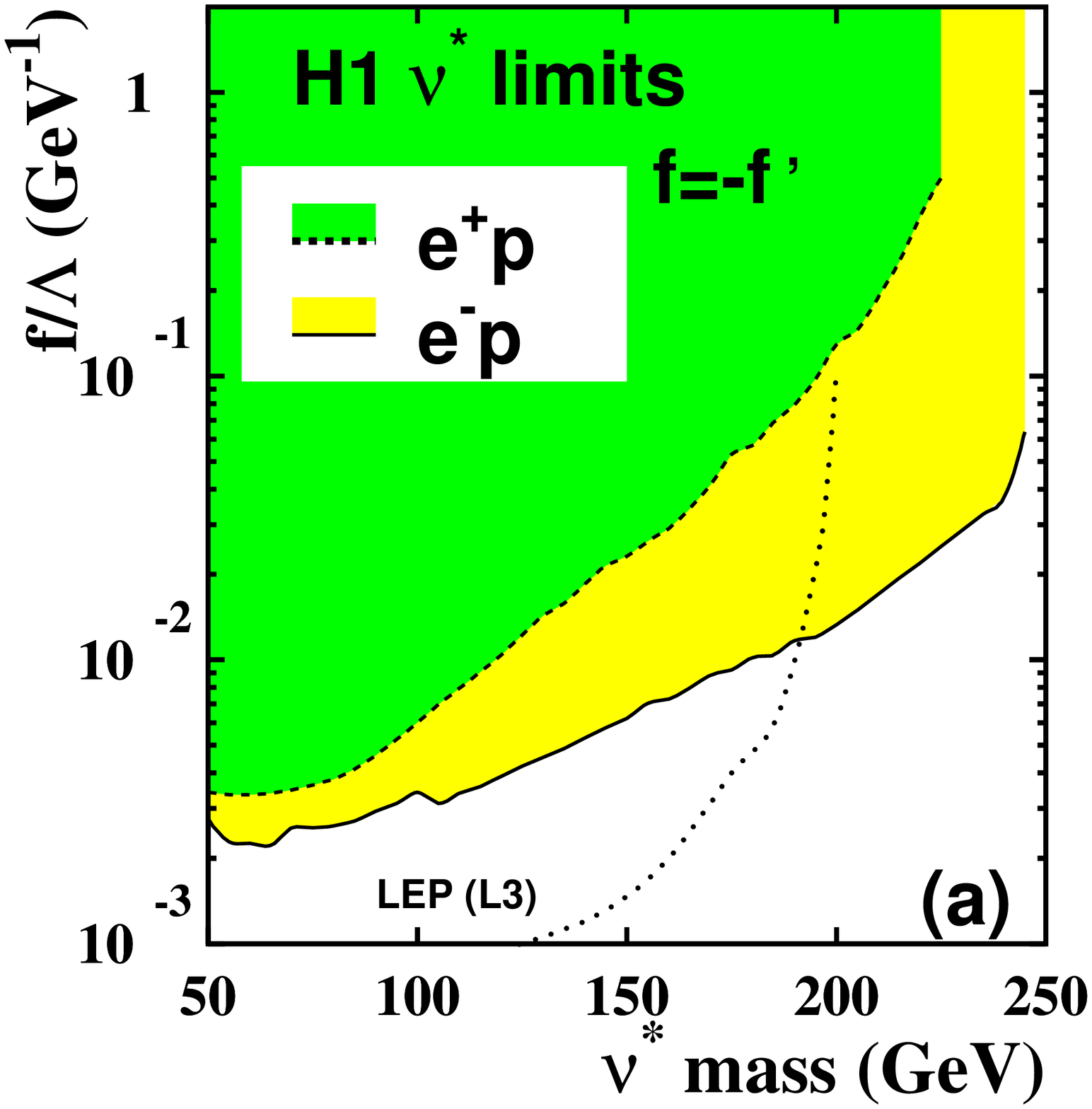,width=7.9cm,height=8.cm}
  \psfig{file=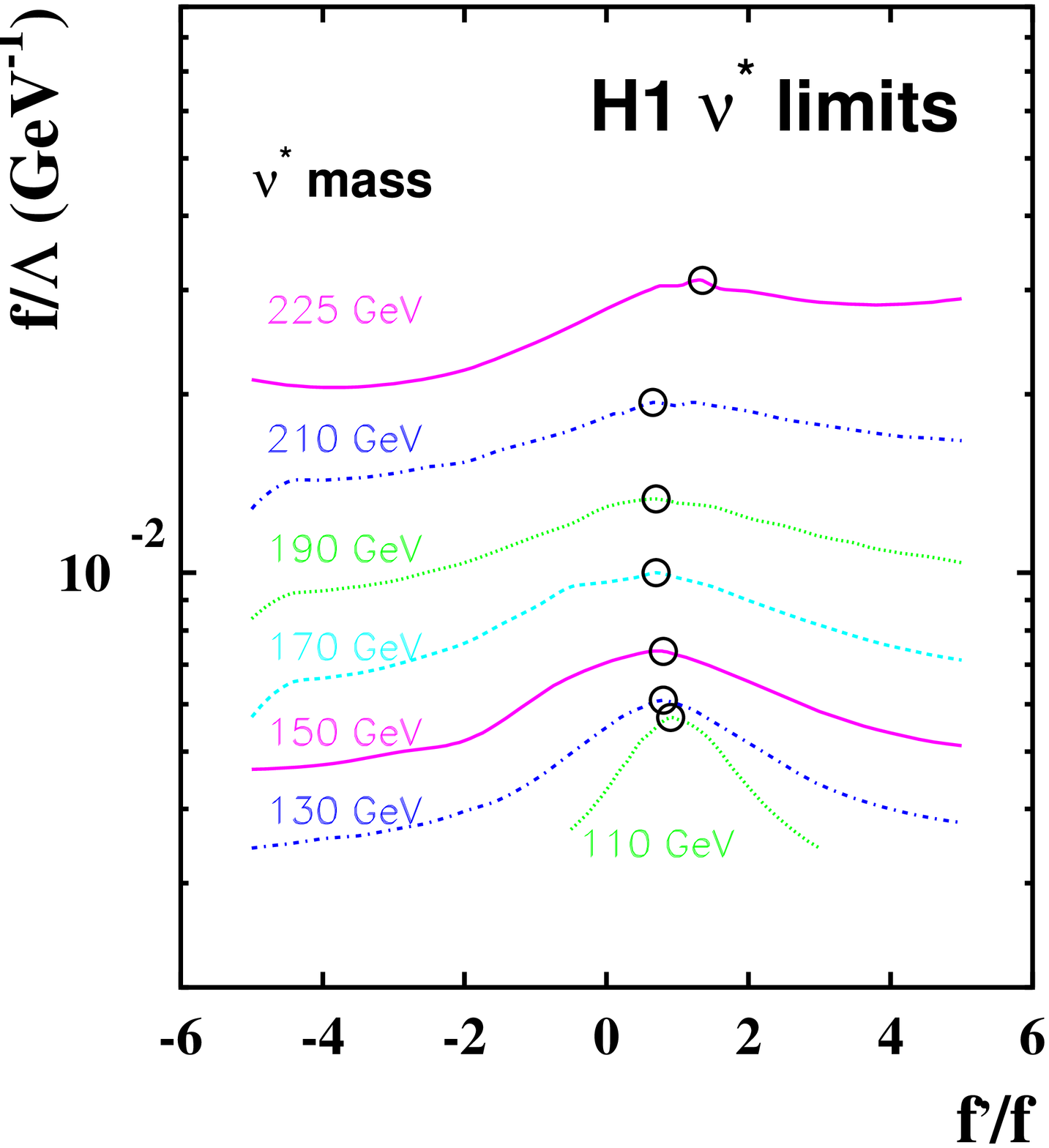,width=7.9cm,height=8.cm}}
  \caption{Left: limits at $95\%$ C.L.\ on $f/\Lambda$ from the H1
           search for excited neutrinos compared to a corresponding
           result from L3. Right: dependence of the H1 limits on the
           ratio $f'/f$.}
  \label{fig-ns-flam}
\end{figure}

The HZK model is used to relate these limits to exclusion plots of $f/\Lambda$
vs.\ $M_{f^\ast}$, where $\Lambda$ is the characteristic mass scale and $f$
denotes the coupling at the $f^\ast$-fermion-boson vertex. In the HZK model,
there are three independent couplings, of which that to gluons ($f_s$) is
irrelevant for $e^\ast$ and $\nu^\ast$ production at HERA and those to the U(1)
($f'$) and SU(2) ($f$) gauge fields are either assumed to be related by $f=\pm
f'$, or their ratio is varied in the range $|f'/f|\le5$ (H1). The ratio $|f'/f|$
fixes the relative contributions of photon and $Z$ exchange in $e^\ast$
production and also the branching ratios of the $f^\ast$ decays for a given
$M_{f^\ast}$. The resulting exclusion limits on $f/\Lambda$ for $e^\ast$ are
compared to the corresponding LEP results \cite{lepexo} in \fig{es-flam}. Also
shown is the dependence of these limits on $f'/f$. For $f=-f'$, the $e^\ast$ has
no electromagnetic interactions and hence must be produced by $Z$ exchange and
in addition does not decay to $e\gamma$, thus reducing the HERA sensitivity
substantially. \Fig{ns-flam} shows the exclusion limits for $\nu^\ast$ from the
H1 analysis, together with a recent L3 result \cite{acc-01-01}; similar results
are obtained by ZEUS (not shown). The dependence on $f'/f$ is weaker than in the
$e^\ast$ case since $\nu^\ast$ production requires the exchange of a $W$
boson. Minimal sensitivity is obtained around $f=f'$, where the decay
$\nu^\ast\to\nu\gamma$ is prohibited.

Assuming $f/\Lambda=1/M_{f^\ast}$, the results exclude at $95\%$~C.L.\ the
existence of $e^\ast$ with masses between $100$ and about $255\gev$ for $f'=f$
(estimated from \fig{es-flam}), and of $\nu^\ast$ between $100$ and
$135(158)\gev$ for $f'=-f$ ($f'=f$) \cite{zeus-01-08}.

The HERA limits shown in \figand{es-flam}{ns-flam} are weaker than the LEP
limits for $f^\ast$ masses below about $200\gev$, where $e^\ast$ and $\nu^\ast$
could be directly produced in reactions of the type $e^+e^-\to ff^\ast$. For
higher $e^\ast$ masses, the LEP sensitivity comes from cross section deviations
from the SM expectation for the reaction $e^+e^-\to\gamma\gamma$ which would be
induced by $t$-channel exchange of an $e^\ast$; in this region, the HERA and LEP
limits are of similar magnitude, indicating substantial discovery potential in
future high-luminosity HERA running.

\section{Direct searches for leptoquarks and $R_P$-violating squarks}
\label{sec-lq}

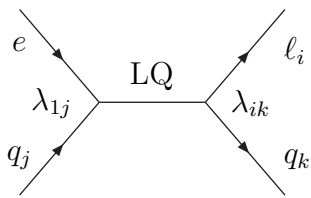
\begin{figure}[b]
  \sidecaption
  \begin{picture}(125.,70.)(0.,0.)
  \SetOffset(0.,-30.)
  \ArrowLine(10,30)(40,65)           
  \ArrowLine(10,100)(40,65)          
  \Line(40,65)(80,65)                
  \ArrowLine(80,65)(110,30)          
  \ArrowLine(80,65)(110,100)         
  \Text(10,90)[t]{$e$}              
  \Text(110,90)[lt]{$\ell_i$}       
  \Text(10,40)[b]{$q_j$}             
  \Text(110,40)[lb]{$q_k$}           
  \Text(60,70)[b]{LQ}                
  \Text(30,65)[rc]{$\lambda_{1j}$}   
  \Text(90,65)[lc]{$\lambda_{ik}$}   
  \end{picture}
  \caption{Feynman graph of the resonant production of an $F\eql2$ LQ in
           electron-quark scattering. The subscripts of the Yukawa couplings at
           the production and decay vertices of the LQ indicate the generations
           of the leptons and quarks involved.}  \label{fig-lq-fey}
\end{figure}

Leptoquarks (LQs) are hypothesized bosons coupling to lepton-quark pairs.  They
could be resonantly produced in $ep$ reactions according to the diagram of
\fig{lq-fey}.  Buchm\"uller, R\"uckl and Wyler (BRW) \cite{buc-87-01} have
classified LQs which (i) conserve the SM gauge symmetries, (ii) only couple to
quarks, leptons and SM gauge bosons and (iii) only have flavor-diagonal
couplings, in ten different states characterized by their fermion number
($|F|\eql2$ or $F\eql0$), spin (scalar ($S$) or vector ($V$)) and weak isospin.
Premise (ii) implies that LQs only decay into $\ell^\pm q$ or $\nu q'$ pairs
with given branching ratios. In many recent LQ searches, this assumption is
dropped and a variable branching ratio $\beta_\ell=\BR(\LQ\to\ell^\pm q)$ is
assigned to the LQs (accounting e.g.\ for the case of $R_P$-violating squarks).

\subsection{First-generation leptoquarks}
\label{sec-lq-fg}

\begin{figure}[b]
  \psfig{file=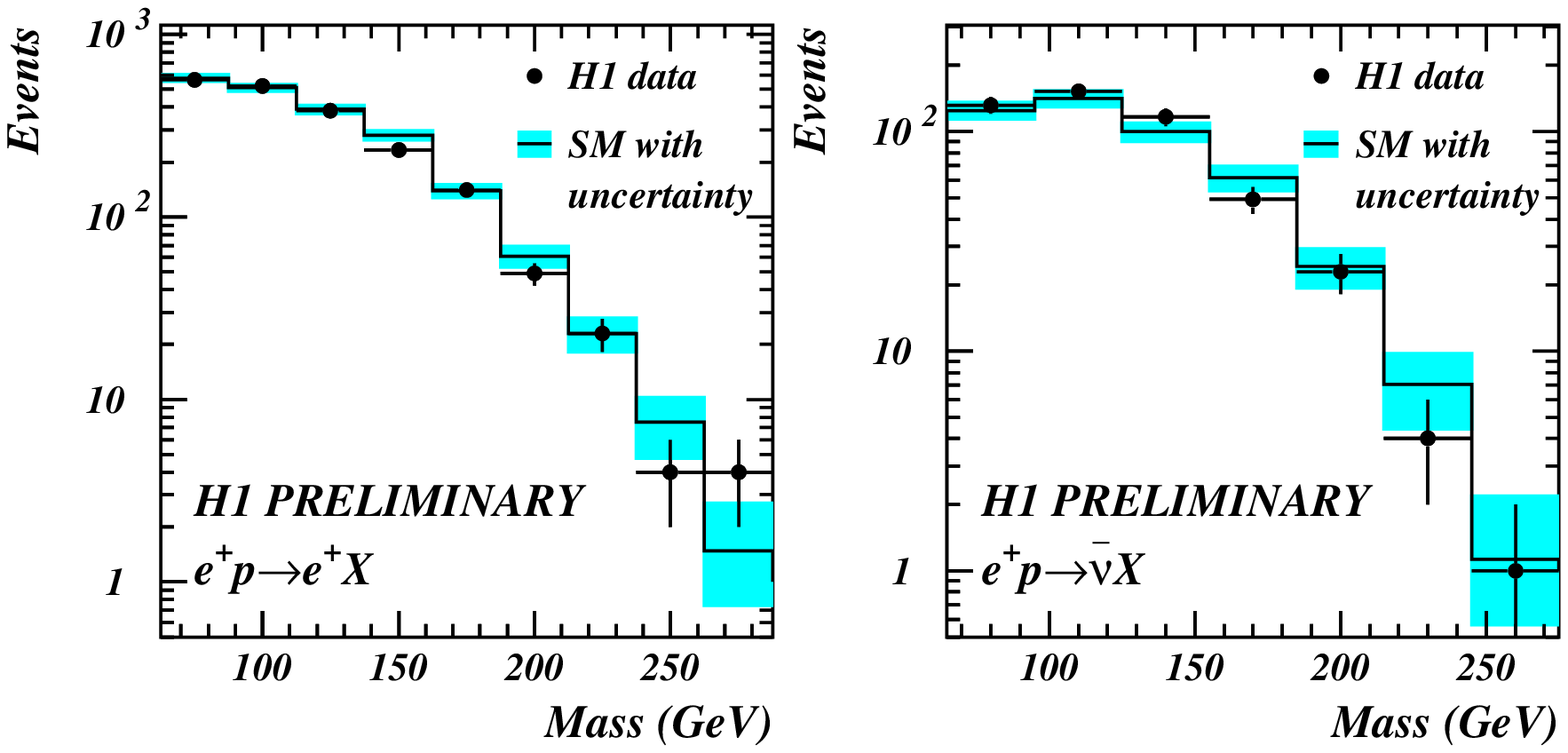,width=\textwidth,height=6.5cm}
  \centerline{
  \psfig{file=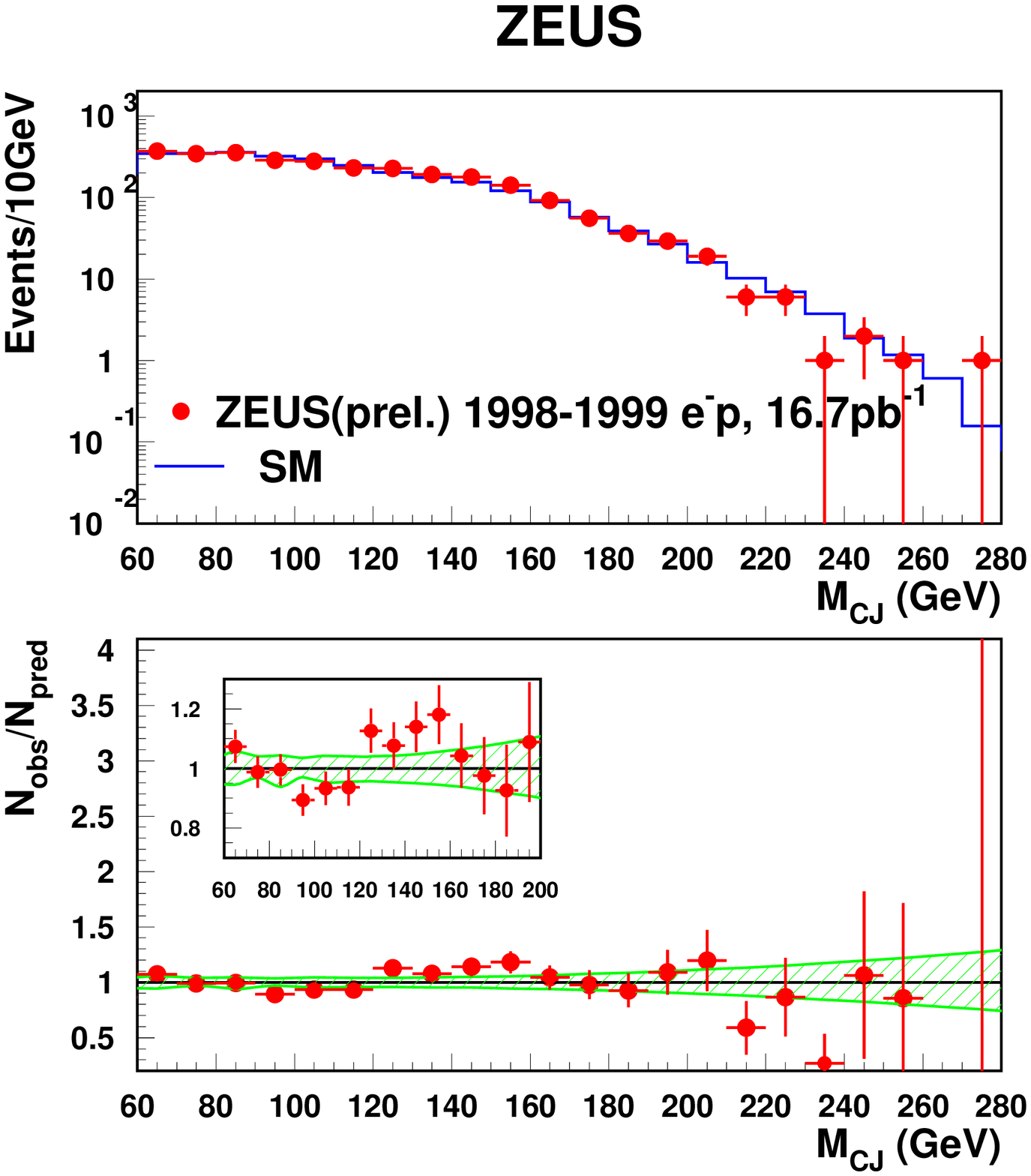,width=7.7cm,height=9.5cm}\hfill
  \psfig{file=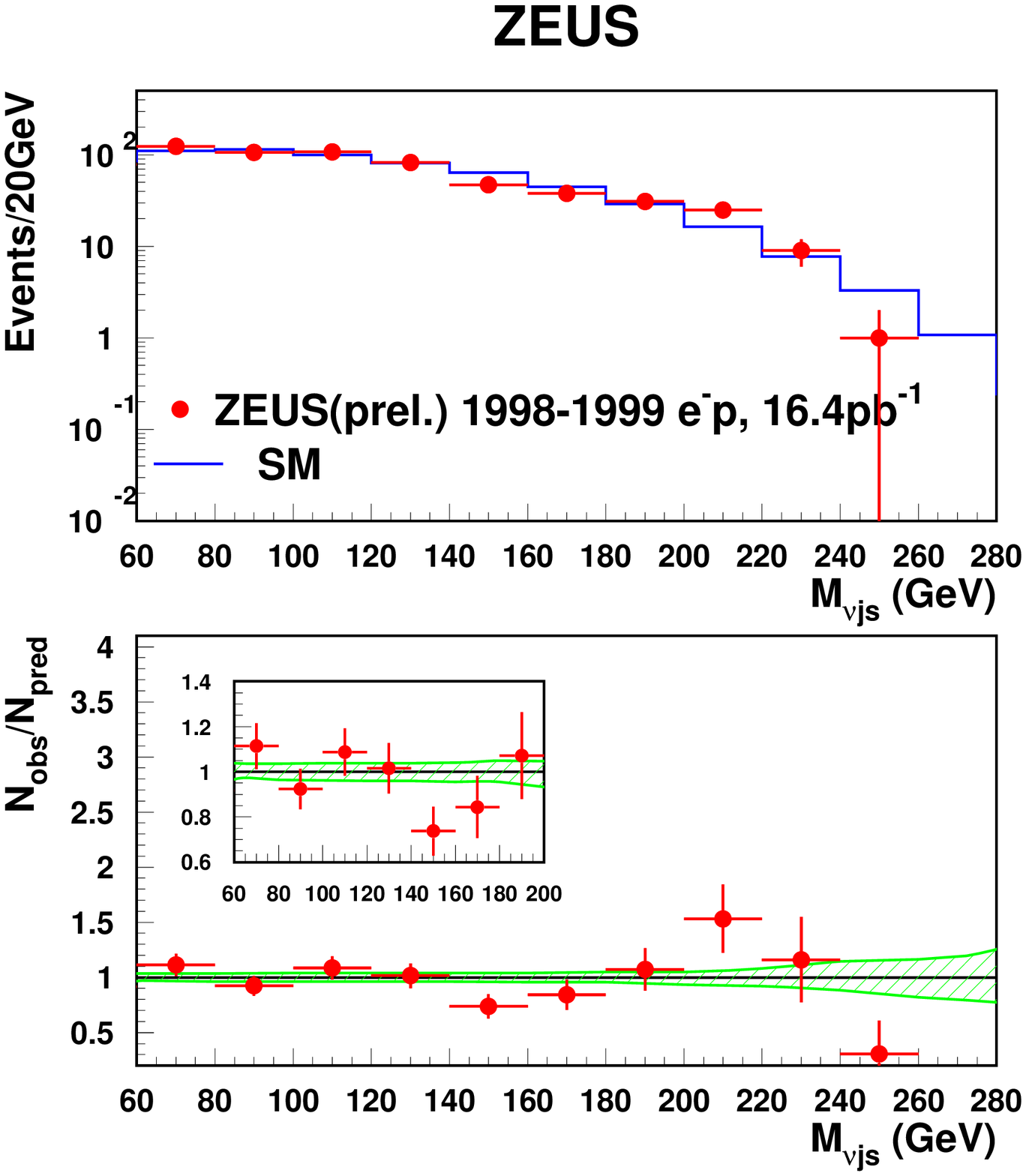,width=7.7cm,height=9.5cm}}
  \caption{Distributions of the reconstructed mass $M=\sqrt{xs}$ for the H1
           $e^+p$ data (top \pcite{h1-prelim-02-064}) and the ZEUS $e^-p$ data
           (bottom \pcite{zeus-amsterdam-907}).  The left column shows the NC,
           the right column the CC data sets.  The symbols with statistical
           error bars represent the data, the histograms the DIS simulation.
           The shaded areas indicate the uncertainties of the SM predictions.}
           \label{fig-lq-mdis}
\end{figure}

LQs coupling only to first-generation leptons can be produced at HERA in
reactions of the type $ep\to\LQ+X\to eq+X\;(\nu_e q'+X)$, yielding the same
event topologies as NC (CC) DIS at high $eq$ invariant mass. The distinctive LQ
signature is a resonant peak in the $\sqrt{xs}$ distribution around the LQ mass,
$M_\LQsub$, where $\sqrt{s}$ is the $ep$ center-of-mass energy and the Bjorken
variable, $x$, is the fraction of the proton momentum carried by the struck
quark.  Furthermore, the LQ decay angular distribution implies harder
distributions in $y=s/(xQ^2)$ than for DIS ($Q^2$ is the negative square of the
four-momentum transfer between $e$ and $p$). LQ production at HERA received
particular attention after H1 \cite{h1-97-06} and ZEUS \cite{zeus-97-03}
reported an excess of events of NC DIS at high $x$ and $Q^2$ in their 1994--1996
data.

\begin{figure}[b]
  \centerline{
  \psfig{file=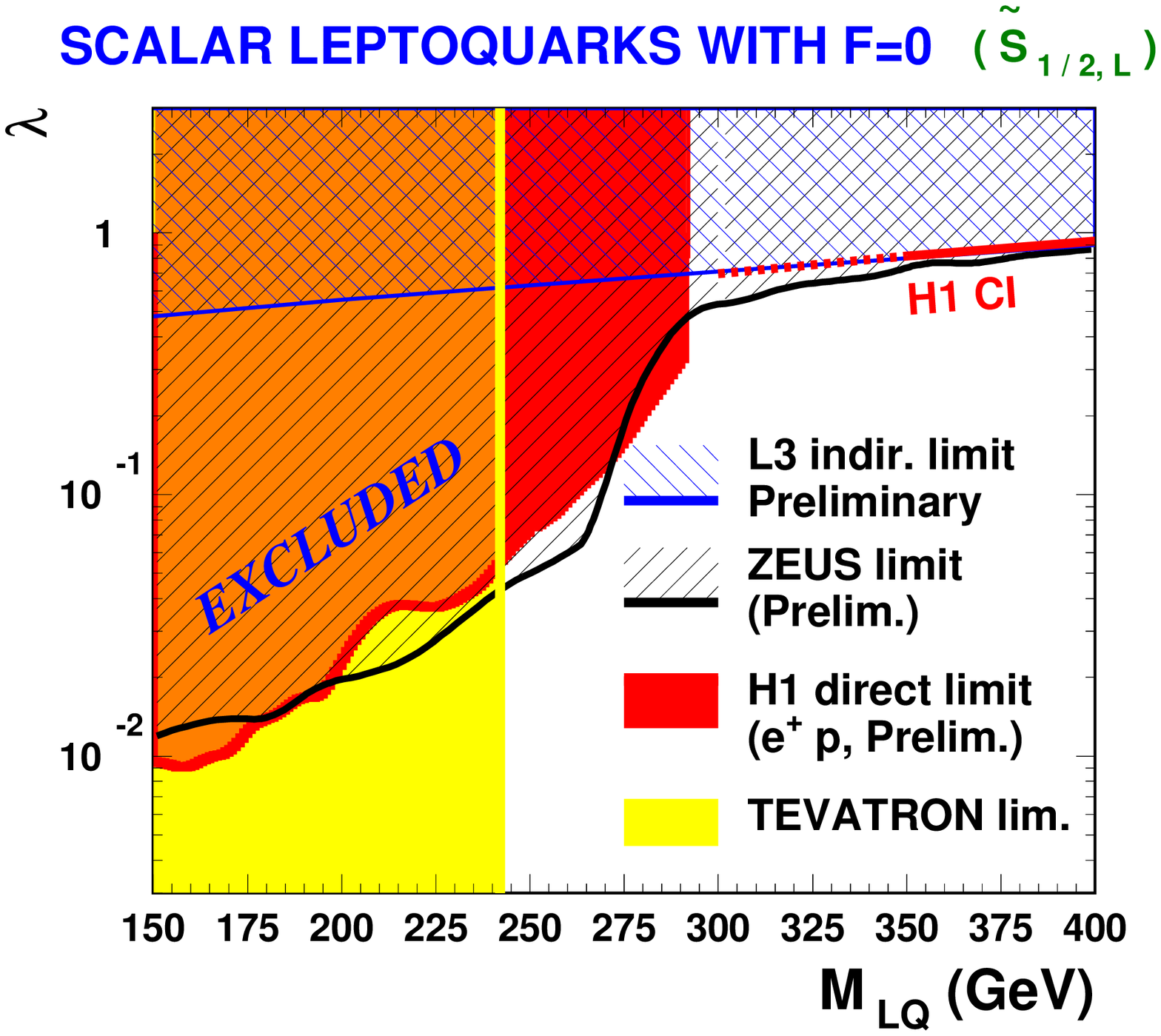,width=7.9cm}\hfill
  \psfig{file=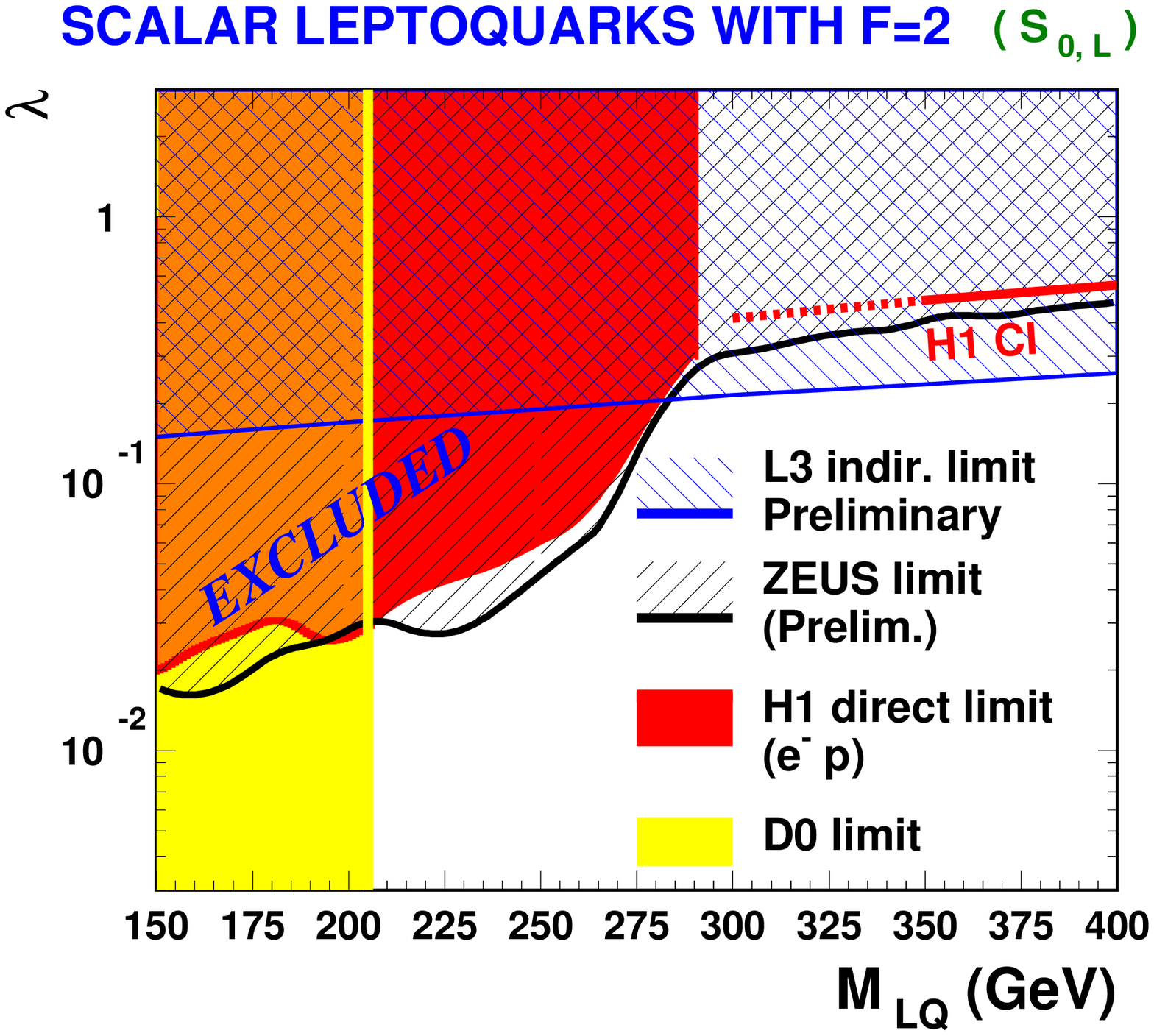,width=7.7cm}}
  \caption{ZEUS and H1 exclusion limits at $95\%$ C.L.\ on the LQ Yukawa
           coupling $\lambda=\lambda_{11}$ as a function of the LQ mass for two
           representative scalar LQ species, $\tilde S_{1/2,L}$ (left) and
           $S_{0,L}$ (right). Also shown are the results from indirect searches
           at LEP (L3) and the search for LQ pair production at the Tevatron.}
           \label{fig-lq-lim}
  \vspace*{3.mm}

  \centerline{
  \psfig{file=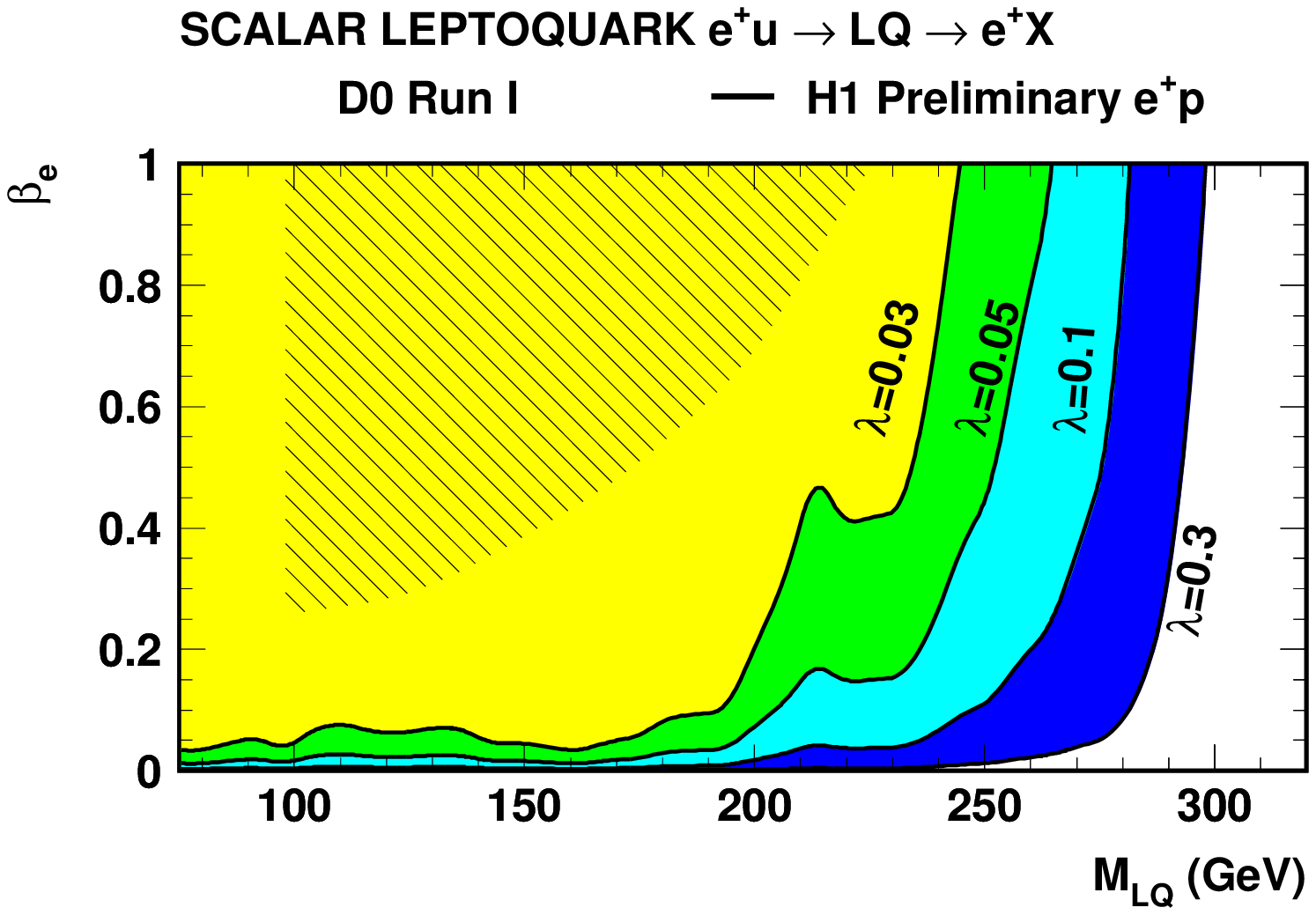,width=7.6cm}\hfill
  \psfig{file=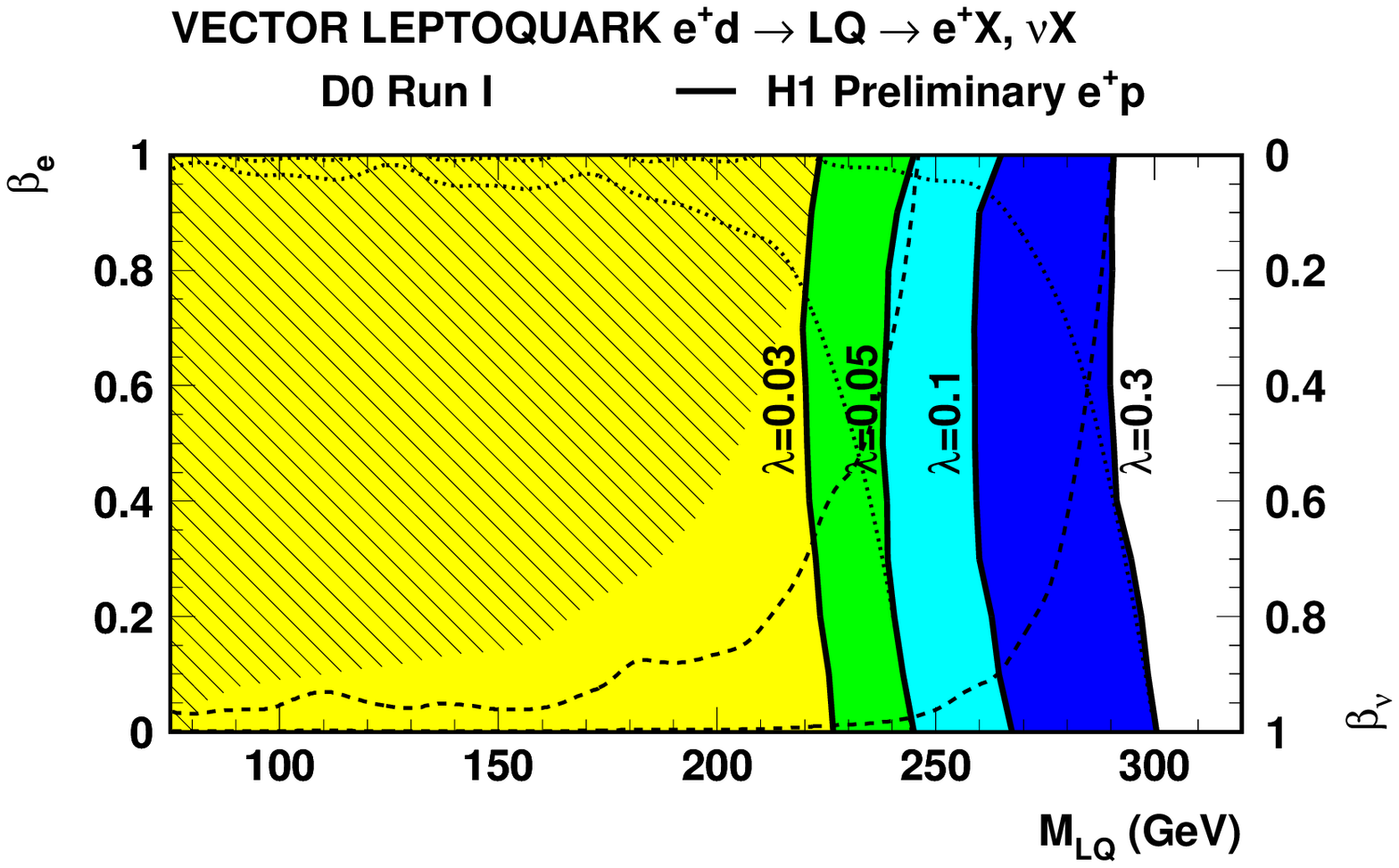,width=8.0cm}}
  \caption{H1 exclusion limits on $\beta_e$ as a function of $M_\LQsub$ for
           different assumptions on $\lambda$, for a scalar LQ coupling to
           $e^+u$ (left) and for a vector LQ coupling both to $e^+d$ and $\nubar
           u$ (right). In the right plot, the exclusion limits obtained from
           the NC and CC channels alone are indicated by dashed and dotted
           lines, respectively. The hatched areas denote the regions excluded by
           \DO \pcite{abb-97-01,*abb-97-02}.}
  \label{fig-lq-h1beta}
\end{figure}

ZEUS \cite{zeus-amsterdam-907} and H1 \cite{h1-01-02,h1-prelim-02-064} have
searched in the available DIS data for deviations from the SM prediction
consistent with a LQ signal. The data agree well with the SM and
show no sign of LQ production. The previously reported event excess has not been
confirmed by the new high-statistics data.  As representative examples, the
distributions of the reconstructed mass $M=\sqrt{xs}$ are shown in
\fig{lq-mdis} for the H1 $e^+p$ data (top) and the ZEUS $e^-p$ data (bottom).
In the NC channel ($ep\to eX$), H1 derives $M$ from the energy and angle of the
scattered electron, whereas ZEUS uses the electron energy and the kinematic
variables of hadronic jets with transverse momentum exceeding $15\gev$. For CC
reactions ($ep\to\nuan X$), H1 reconstructs $M$ from the energy, longitudinal
and transverse momentum of the hadronic final state, whereas ZEUS derives $M$ as
the invariant mass of the reconstructed neutrino and the jets with a transverse
momentum above $10\gev$.

\looseness=-1
In the absence of a LQ signal, exclusion limits on the Yukawa coupling,
$\lambda\eql\lambda_{11}$, are derived as a function of $M_\LQsub$ for LQs
coupling to first-generation quarks and leptons ($i=j=k=1$ in \fig{lq-fey}).
This analysis is based on the BRW model.  The results are shown in
\fig{lq-lim} for two representative scalar LQ species with $|F|\eql2$ and $F\eql0$,
respectively. In order to set limits, the event distributions in $M$ and $y$ are
compared to MC simulations of DIS reactions and of signal events. For LQs that
can decay both to $eq$ and $\nu q'$, the limits are inferred from the NC and CC
data sets. Taking into account $u$-channel LQ exchange and the DIS-LQ
interference, in addition to the $s$-channel LQ formation of \fig{lq-fey} (which
dominates for $M_\LQsub\lesssim240\gev$), yields sensitivity to LQ masses up to
and beyond the HERA center-of-mass energy (ZEUS). Alternatively, a contact
interaction approach (see \sec{ind}) can be used to constrain high-mass LQs
(H1).  Note that the sensitivity is highest for LQs that can be produced from
valence quarks, i.e.\ for LQs with $F\eql0$ in $e^+p$ and for LQs with $F\eql2$
in $e^-p$ scattering.  Also indicated in \fig{lq-lim} are the exclusion limits
obtained by L3 \cite{l3-amsterdam-462} (from a search for anomalous
contributions to the cross section $\sigma(e^+e^-\to\text{hadrons})$ induced by
the $t$-channel exchange of LQs) and from the Tevatron, where LQs could be
pair-produced with a rate that is independent of $\lambda$.  The combined CDF
and \DO limit is $M_\LQ\gre242\gev$ \cite{gro-98-01} for scalar LQs with
$\beta_e\eql1$ and is expected to be even higher for vector LQs.

H1 has derived LQ exclusion limits as functions of $\beta_e$ and $M_\LQ$ for
fixed values of $\lambda$ \cite{h1-prelim-02-064} (\fig{lq-h1beta}), thus
allowing a direct comparison to the LQ exclusion region from \DO
\cite{abb-97-01,*abb-97-02}. \Figand{lq-lim}{lq-h1beta} demonstrate that
the HERA experiments have a unique discovery potential for LQs with low
$\beta_e$ and masses beyond about $200\gev$.

\subsection{$R_P$-violating squarks}
\label{sec-lq-rps}

In supersymmetric models without conservation of $R$-parity,
$R_P=(-1)^{3B+L+2S}$ ($B$ being the baryon number, $L$ the lepton number and $S$
the spin), squarks can have Yukawa couplings $\lambda'_{ijk}$ to leptons and
quarks and thus behave like LQs (see \cite{dre-97-01,stmp-168} and references
therein).  The generation indices are $i=1$ for the electron, $j$ for the squark
and $k$ for the quark. The dominant production cross section, and thus the
highest sensitivity at HERA, is expected for $\tilde u$-type squarks which can be
produced from $d$ quarks, whereas all other squarks couple only to the quark sea
in the proton. The searches at HERA focus on this scenario, assuming that one
coupling $\lambda'_{1j1}$ is non-zero and other flavor combinations do not
contribute to the cross section.

\begin{figure}[t]
  \begin{picture}(440.,140.)(0.,-25.)
  \SetWidth{0.75}
  \SetOffset(0.,0.)
  \ArrowLine(10,30)(40,70)           
  \ArrowLine(10,110)(40,70)          
  \Line(40,70)(80,70)                
  \ArrowLine(80,70)(110,110)         
  \DashLine(80,70)(110,40){4}        
  \ArrowLine(110,40)(140,10)         
  \ArrowLine(140,70)(110,40)         
  \ArrowLine(140,10)(170,40)         
  \ArrowLine(170,-20)(140,10)        
  \CCirc(40,70){2}{Black}{Black}
  \CCirc(140,10){2}{Black}{Black}  
  \Text(10,102)[rt]{$e^+$}           
  \Text(8,35)[b]{$q^k$}              
  \Text(115,105)[lb]{$q$}            
  \Text(60,77)[b]{$\tilde q^j$}      
  \Text(30,70)[rc]{$\lsqp{1jk}$}     
  \Text(88,50)[rc]{$\chi^0$}         
  \Text(118,20)[rc]{$\tilde q,\tilde e$}
  \Text(145,70)[lc]{$\qbar,e$}       
  \Text(175,-20)[lc]{$e,\qbar''$}    
  \Text(175,40)[lc]{$q'$}            
  \Text(150,10)[lc]{$\lsqp{1jk}$}    
  \SetOffset(240.,0.)
  \ArrowLine(10,30)(40,70)           
  \ArrowLine(10,110)(40,70)          
  \Line(40,70)(80,70)                
  \ArrowLine(80,70)(110,110)         
  \DashLine(80,70)(110,40){4}        
  \DashLine(110,40)(140,70){3}       
  \Photon(110,40)(140,10){4}{3}      
  \ArrowLine(140,10)(170,40)         
  \ArrowLine(170,-20)(140,10)        
  \CCirc(40,70){2}{Black}{Black}
  \CCirc(140,70){2}{Black}{Black}
  \Text(10,102)[rt]{$e^+$}           
  \Text(8,35)[b]{$q^k$}              
  \Text(115,105)[lb]{$q$}            
  \Text(60,77)[b]{$\tilde q^j$}      
  \Text(30,70)[rc]{$\lsqp{1jk}$}     
  \Text(88,50)[rc]{$\chi^\pm$}       
  \Text(118,20)[rc]{$W^\pm$}         
  \Text(150,75)[cb]{$\chi^0\to eq'\qbar$}  
  \Text(175,-20)[lc]{$\qbar,\nu$}    
  \Text(175,40)[lc]{$q',\ell$}       
  \end{picture}
  \caption{Examples for cascade decays of squarks in $R_P$-violating
           supersymmetry. The $R_P$-violating vertices are marked with
           dots. The symbols $\chi^0$ and $\chi^\pm$ denote the neutralino
           and the chargino, respectively.}
  \label{fig-rp-fey}
\end{figure}
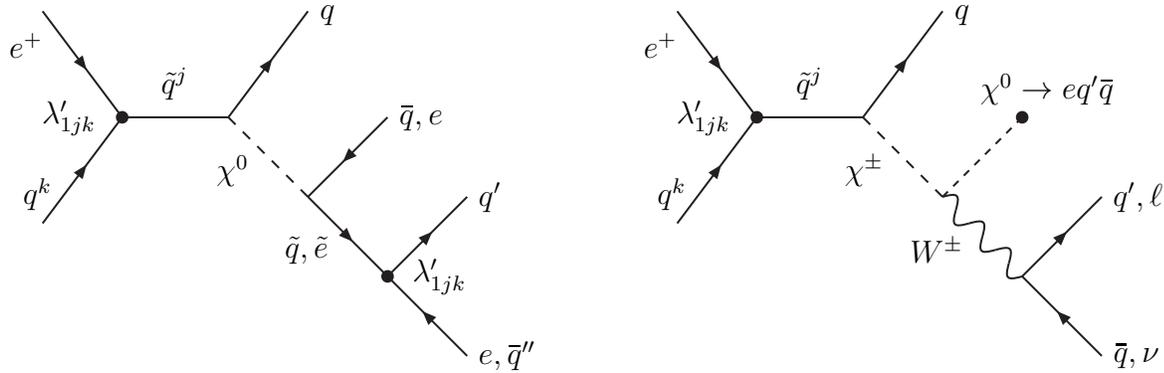

\begin{figure}[t]
  \centerline{
  \psfig{file=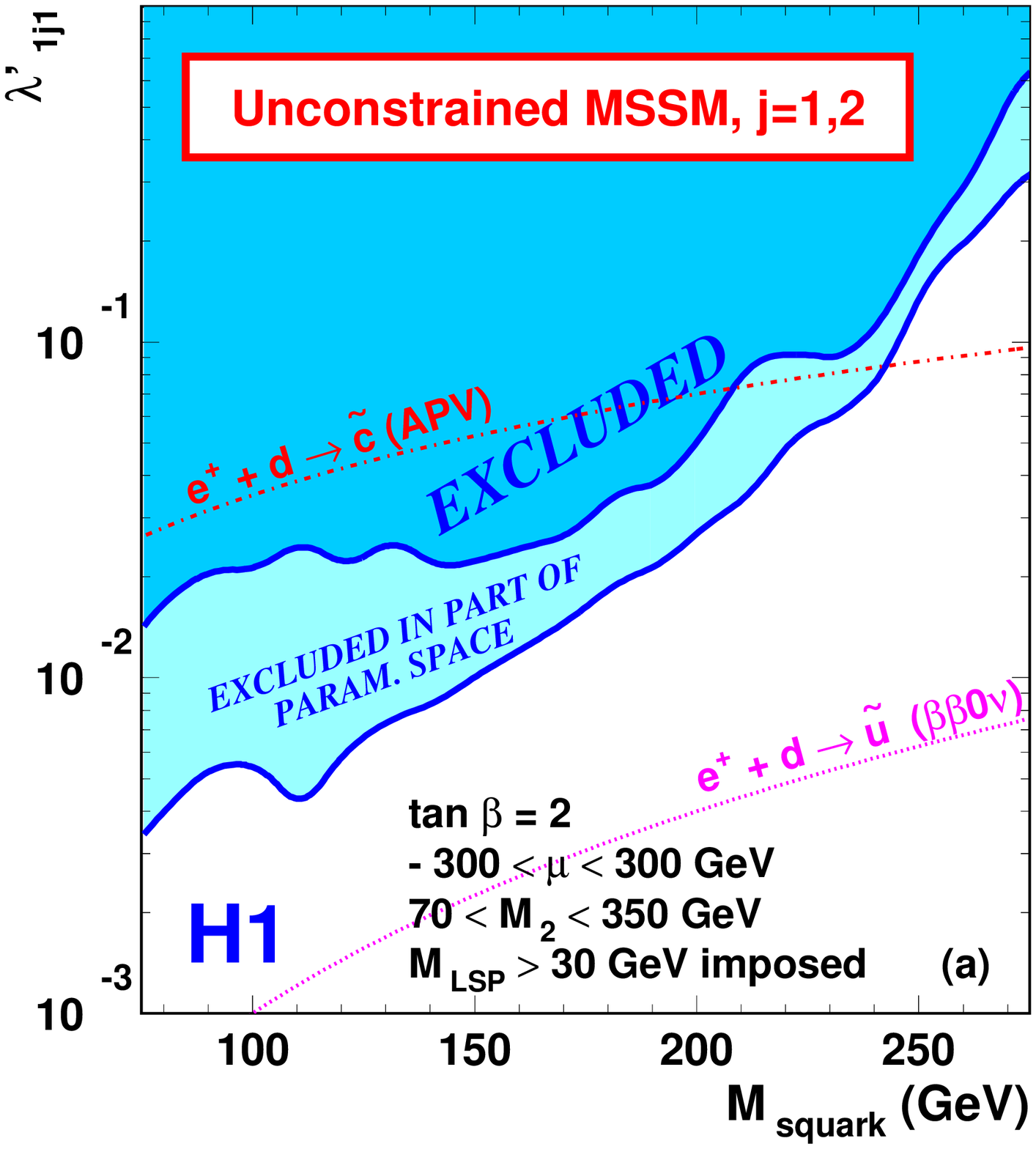,width=7.4cm}\hfill
  \psfig{file=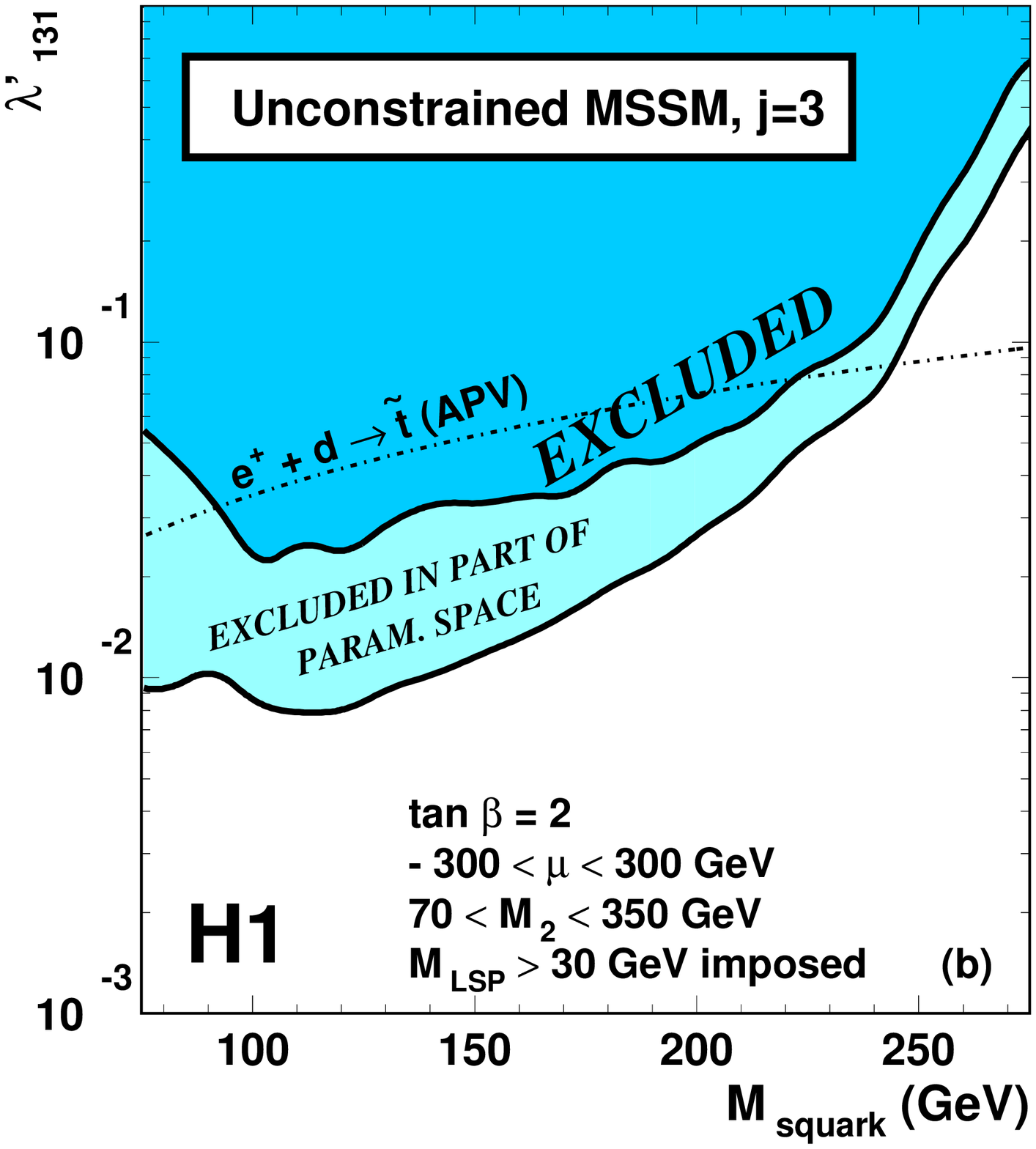,width=7.4cm}}
  \caption{H1 exclusion limits on $\lambda'_{1j1}$ as a function of the squark
           mass for $\tilde u$ and $\tilde c$ squarks ($j=1,2$, left) and for
           $\tilde t$ squarks ($j=3$, right). For fixed $\tan\beta$, the
           supersymmetric parameters $\mu$ and $M_2$ have been scanned in the
           ranges indicated on the plots. The dark shaded region is excluded for
           all parameter combinations, whereas the light shaded region is
           excluded only in part of the parameter space. The dotted and
           dash-dotted lines indicate the exclusion limits from low-energy
           measurements \pcite{dre-97-01}.}
  \label{fig-rp-lim}
\end{figure}

In addition to the LQ-like decay $\tilde q_j\to ed$, squarks can also undergo
cascade gauge decays, with $R_P$-violating processes occuring further down the
chain. Examples for such cascades are shown in \fig{rp-fey}. These reactions
produce events with distinctive signatures: several jets and one or more leptons
with high transverse momentum, where in some channels the final-state electron
can have opposite charge to that of the beam electron. Using about $40\pbi$ of
$e^+p$ data collected in 1994--97, ZEUS \cite{zeus-osaka-1042} and H1
\cite{h1-01-01} have searched for squark production in $e^+p$ scattering by
combining the results of the LQ analyses in the NC channel (squarks cannot decay
to $\nuan q$) with dedicated searches for events that are consistent with squark
cascade decays. No significant excess over the SM background was found, and
exclusion limits on $\lambda'_{1j1}$ as a function of the squark mass have been
set.  The H1 results are shown in \fig{rp-lim}. It can be seen that for $j=2,3$
the sensitivity of HERA greatly exceeds that of the currently most sensitive
low-energy probe, atomic parity violation (APV); for $j=1$, however, the limits
obtained from neutrinoless double-beta decay are much stronger than those from
HERA \cite{dre-97-01}.

\subsection{Lepton-flavor violating leptoquarks}
\label{sec-lq-lfv}

\begin{figure}[t]
  \psfig{file=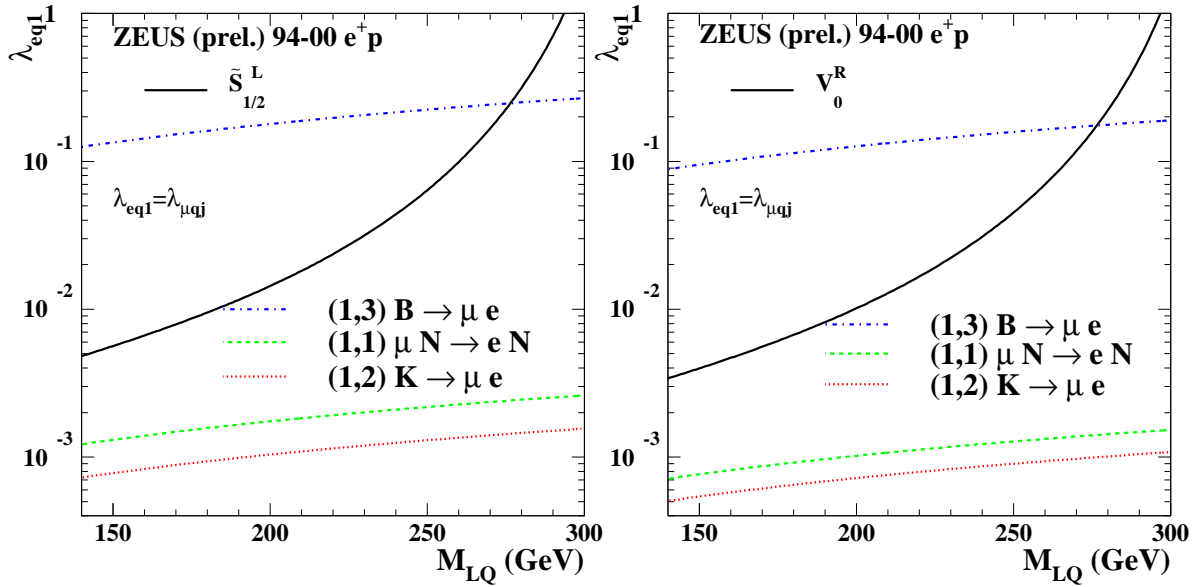,width=\textwidth}
  \caption{ZEUS exclusion limits at $95\%$ C.L.\ on the coupling strength
           $\lambda_{2k}$ of LFV LQs under the assumption
           $\lambda_{2k}=\lambda_{11}$. The limits are shown for scalar (left)
           and vector (right) LQ species that couple to $d$ quarks and thus
           yield minimal sensitivity in the ZEUS search. Also shown are the
           limits from rare processes \pcite{dav-94-01,*gab-00-01,*pdg-02-01},
           where the numbers in parentheses indicate the quark generations that
           couple to the $e$ and the $\mu$, respectively.}
  \label{fig-lfv-lim}
\end{figure}

Leptoquarks or squarks that couple to leptons of several generations
simultaneously (i.e.\ $\lambda_{1j}\ne0$ and $\lambda_{ik}\ne0$ for at least one
flavor combination with $i>1$, see \fig{lq-fey}) could induce lepton-flavor
violating (LFV) DIS-like reactions at HERA with the final-state $e$ replaced by
a $\mu$ or $\tau$.  Searches for such events in the 1994--97 $e^+p$ data by ZEUS
\cite{zeus-01-12} and H1 \cite{h1-99-03}, as well as an update by ZEUS for the
$\mu$ channel based on the full $e^+p$ data sample \cite{zeus-amsterdam-906},
yielded no evidence for LFV.  The exclusion limits \cite{zeus-amsterdam-906} for
the LQ couplings resulting under the assumption $\lambda_{11}=\lambda_{2k}$
(i.e.\ $\BR(\LQ\to\mu q_k)=1/2$) are shown in \fig{lfv-lim}. Assuming couplings
of electromagnetic strength, $\lambda_{11}=\lambda_{2k}=\sqrt{4\pi\alpha}$,
these limits correspond to $M_\LQsub>279$--$301\gev$, depending on the LQ
type. Comparison of the ZEUS results with limits derived from searches for
low-energy rare processes \cite{dav-94-01,*gab-00-01,*pdg-02-01} (see
\fig{lfv-lim}) shows that the HERA data are highly sensitive to LFV
LQs coupling to 3rd-generation quarks but less so if $k=1$ or if $q_k$ is a $s$
quark. The Tevatron experiments report mass limits on LQs decaying to $\mu q$ or
$\tau q$ (see \cite{kuz-02-01} and references therein), but since LQs would be
pair-produced in $p\pbar$ reactions via QCD gauge couplings, these results do
not constrain LFV.

\section{Indirect leptoquark and squark searches}
\label{sec-ind}

\begin{figure}[t]
  \sidecaption
  \psfig{file=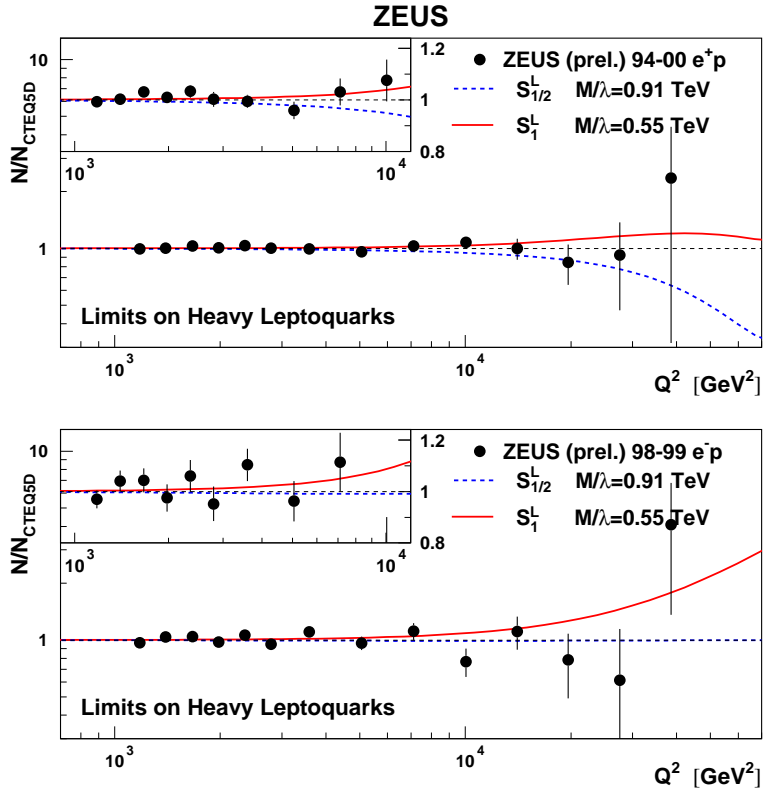,width=10.cm}
  \caption{Ratios of measured to predicted $Q^2$ distributions in the ZEUS $e^+p$
           (top) and $e^-p$ (bottom) NC data.  The solid (dotted) lines show the
           expected modifications induced by $S_1^L$ ($S_{1/2}^L$) LQs with
           $M_\LQsub/\lambda$ corresponding to the $95\%$ C.L.\ lower limit.}
  \label{fig-ci-dist}
\end{figure}

As mentioned in \sec{lq-fg}, LQs and squarks coupling to electrons and quarks
affect the $ep$ DIS cross section through virtual $u$- and $s$-channel exchange,
even if their masses exceed the HERA center-of-mass energy; note that in this
case the characteristic resonance-like signal in $xs$ disappears.  The search
for such high-mass signatures is usually performed using a contact interaction
approach, which parameterizes the cross section modification in terms of a
global mass scale, $\Lambda$, and of sets of chiral couplings of the
hypothesized heavy object to the different quark flavors. The results of such
analyses have been provided by ZEUS \cite{zeus-99-05,zeus-budapest-602} and H1
\cite{h1-00-01,h1-prelim-02-062}. For the special case of LQs conforming to the
BRW model, the contact interaction limits can be converted to lower bounds on
$M_\LQsub/\lambda$. \Fig{ci-dist} shows the ratio of measured to predicted $Q^2$
distributions in the ZEUS $e^+p$ and $e^-p$ NC data together with the expected
modifications induced by scalar LQs; these lines correspond to the $95\%$ C.L.\
lower limit on $M_\LQsub/\lambda$. Note the complementary sensitivity of the
data sets to the different LQ species. The $M_\LQsub/\lambda$ limits for all BRW
LQ species and for $\tilde u$- and $\tilde d$-type $R_P$-violating squarks as
obtained by ZEUS \cite{zeus-budapest-602} and H1 \cite{h1-prelim-02-062} are
summarized in \tab{ci-lq}. The assumptions implicit in the contact interaction
approach restrict the applicability of these results to high-mass LQs, i.e.\ to
$M_\LQsub\gg\sqrt{s}=318\gev$. As is also seen in \fig{lq-lim}, these limits are
weaker than those from the direct search but still restrictive for LQ masses up
to a few $\Tev$.

\begin{table}[t]
  \sidecaption
  \raisebox{2.2cm}{
  \begin{tabular}{l|c|cc||l|c|cc}
    \stru{2.8ex}{0.pt}
    {LQ or $\tilde q$}&$F$&\multicolumn{2}{c||}{$M_\LQsub/\lambda\,[\Gev]$}&
    {LQ or $\tilde q$}&$F$&\multicolumn{2}{c}{$M_\LQsub/\lambda\,[\Gev]$}\\
    \stru{2.5ex}{0.pt}
    &&ZEUS&\ H1&&&ZEUS&\ H1\\
    \hline
    $S_0^L\text{\ or\ }\tilde d             $&2&\cA750&\cA720&
    $V_0^L                                  $&0&\cA690&\cA770\\
    $S_0^R                                  $&2&\cA690&\cA670&
    $V_0^R                                  $&0&\cA580&\cA640\\
    $\tilde S_0^R                           $&2&\cA310&\cA330&
    $\tilde V_0^R                           $&0&  1030&  1000\\
    $S_1^L                                  $&2&\cA550&\cA480&
    $V_1^L                                  $&0&  1420&  1380\\\hline
    $S_{1/2}^L                              $&0&\cA910&\cA870&
    $V_{1/2}^L                              $&2&\cA490&\cA420\\
    $S_{1/2}^R                              $&0&\cA690&\cA370&
    $V_{1/2}^R                              $&2&  1150&\cA940\\
    $\tilde S_{1/2}^L\text{\ or\ }\tilde u  $&0&\cA500&\cA430&
    $\tilde V_{1/2}^L                       $&2&  1260&  1020\\\hline
  \end{tabular}}
  \caption{Lower limits at $95\%$ C.L.\ on $M_\LQsub/\lambda$ for the BRW LQ
           species derived by the contact interaction analyses of ZEUS
           \pcite{zeus-budapest-602} and H1 \pcite{h1-prelim-02-062}. Note that
           with $\lambda\equiv\lambda'_{1j1}$ the limits for the scalar LQs
           $S_0^L$ and $\tilde S_{1/2}^L$ also apply to $\tilde d$- and $\tilde
           u$-type squarks, respectively }
 \label{tab-ci-lq}
\end{table}

High-mass exclusion limits for the combination $\lambda_{1j}\lambda_{ik}/
M_\LQsub^2$ characterizing LFV LQs can be derived in a manner similar to that
used for the first-generation LQs. From the absence of LFV signal events (which
are very similar to those of the direct search) in the $\mu$ channel, ZEUS
\cite{zeus-amsterdam-906} infers upper limit values on
$\lambda_{1j}\lambda_{2k}/M_\LQsub^2$ for the various combinations of $j$, $k$
and LQ species. The results for LQs with $F\eql0$ are shown in \tab{ci-hmlfv},
the limits for $F\eql2$ are of equal magnitude.  Some of these limits are
stronger than those from other measurements. This also applies for similar
results published previously by ZEUS \cite{zeus-01-12} and H1 \cite{h1-99-03}
for the $\tau$ channel.

\begin{table}[t]
  \caption{ZEUS \pcite{zeus-amsterdam-906} $95\%$ C.L.\ upper limits on
           $\lambda_{1\alpha}\lambda_{2\beta}/M_\LQsub^2$ in units of
           $\Tev^{-2}$, for LQs with $F\eql0$ that couple to $eq_\alpha$ and to
           $\mu q_\beta$.  Each cell of the table shows, from top to bottom, the
           reaction or decay yielding the most restrictive low-energy constraint
           \pcite{dav-94-01,*gab-00-01,*pdg-02-01}, the corresponding limit and
           the ZEUS result (enclosed in a box when stronger than the low-energy
           limit). Stars (*) indicate cases involving a top quark.}
  \label{tab-ci-hmlfv}
  \psfig{file=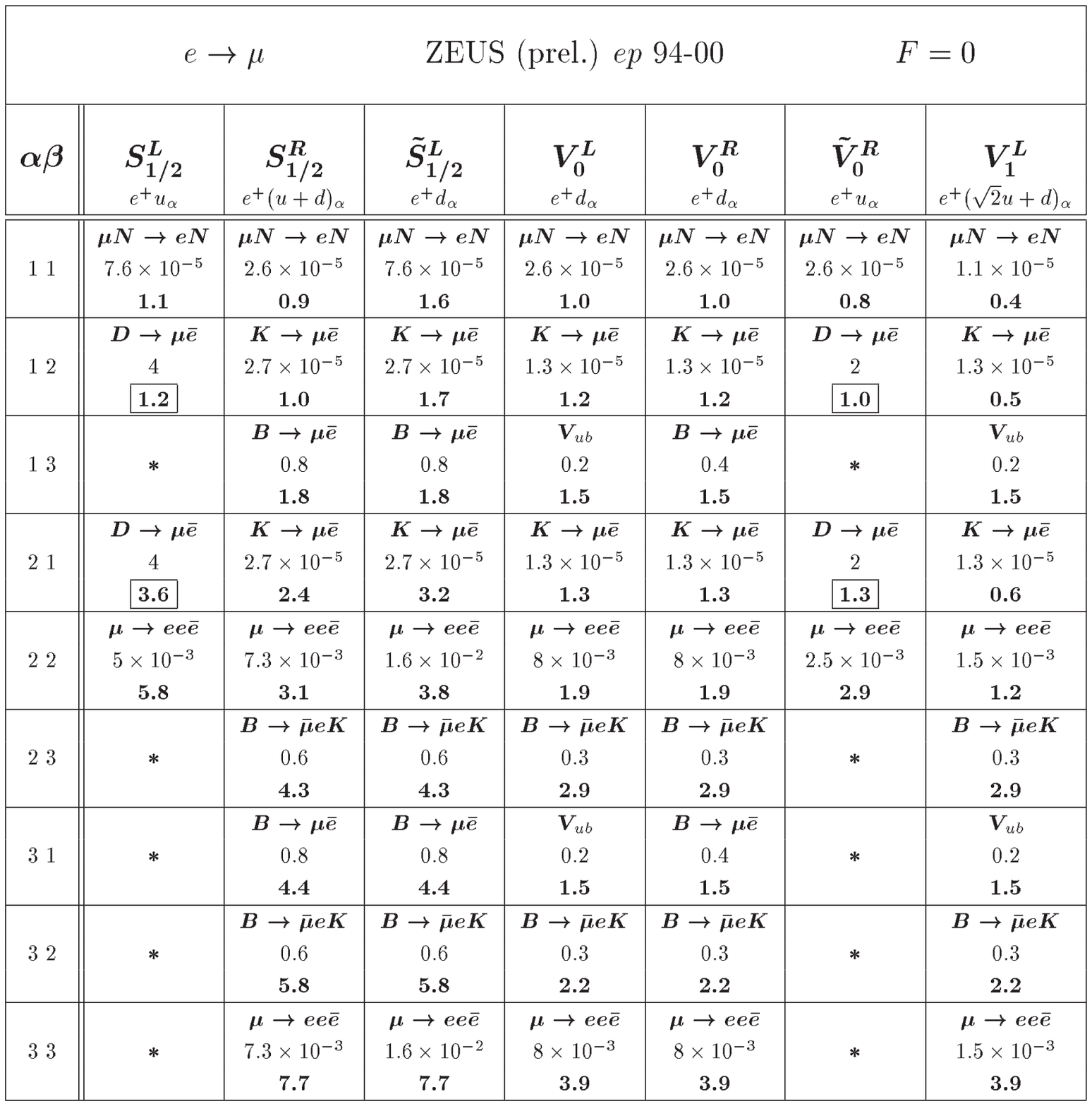,width=\textwidth}
\end{table}

\section*{Acknowledgments} 
None of the research reported here would have been possible without the
dedicated efforts of the HERA crew, and of all those who contributed to the
design, construction, maintenance and operation of the detectors. I'm indebted
to Malcolm Derrick and Tim Greenshaw for the careful reading of the manuscript.


\section*{References}
\renewcommand{\baselinestretch}{0.97}
\bibliographystyle{bib/bsm02}
{
\raggedright
\bibliography{bib/bsm02}
}

\newpage

%
%
\end{document}